\documentclass[twocolumn]{aastex62}

\usepackage{apjfonts}
\usepackage[T1]{fontenc}
\usepackage{color}
\usepackage{amsmath,amstext}
\usepackage{hyperref}
\usepackage{n  atbib}

\newcommand\beq{\begin{equation}}
\newcommand\eeq{\end{equation}}

\shortauthors{Lazzati \& Perna}

\begin{document}

\title{\sc Jet--cocoon outflows from neutron star mergers:  structure, light curves, and fundamental physics}

\author{Davide Lazzati}
\affiliation{Department of Physics, Oregon State University, 301
Weniger Hall, Corvallis, OR 97331, USA}

\author{Rosalba Perna}
\affiliation{Department of Physics and Astronomy, Stony Brook University, Stony Brook, NY, 11794, USA}
\affiliation{Center for Computational Astrophysics, Flatiron Institute, 162 5th Avenue, New York, NY 10010, \
USA}

\begin{abstract}
  The discovery of GW170817, the merger of a binary neutron star (NS)
  triggered by a gravitational wave detection by LIGO and Virgo, has
  opened a new window of exploration in the physics of NSs and their
  cosmological role.  Among the important quantities to measure are
  the mass and velocity of the ejecta produced by the tidally
  disrupted NSs and the delay -- if any -- between the merger and the
  launching of a relativistic jet. These encode information on the
  equation of state of the NS, the nature of the merger remnant, and
  the jet launching mechanism, as well as yielding an estimate of the
  mass available for $r$-process nucleosynthesis.  Here we derive
  analytic estimates for the structure of jets expanding in
  environments with different density, velocity, and radial extent. We
  compute the jet-cocoon structure and the properties of the broadband
  afterglow emission as a function of the ejecta mass, velocity, and
  time delay between merger and launch of the jet. We show that
  modeling of the afterglow light curve can constrain the ejecta
  properties and, in turn, the physics of neutron density matter. Our
  results increase the interpretative power of electromagnetic
  observations by allowing for a direct connection with the merger
  physics.
\end{abstract}


\section{Introduction}

The discovery of GRB170817A in association with the gravitational wave
(GW) source GW170817, produced by the merger of a double neutron star
(NS) \citep{Abbott2017a}, has opened a new era in the study of short
gamma-ray bursts (SGRBs) as well as of NSs, both separately, but
especially when the information from GWs and the electromagnetic
counterparts are combined.  Prior to the association with this GW
event, only the brightest SGRBs were followed at longer
wavelengths. However, even under these conditions, the afterglows were
generally dim, and rarely full broadband coverage was possible
\citep{Fong2015}.

GRB170817, despite having an isotropic luminosity $\sim 10^4$ times
lower than that of 'standard' SGRBs ($L\sim10^{51}$~erg/s), was very
bright due to its relatively small distance of 40~Mpc, having the
initial trigger been in GWs rather than in $\gamma$-rays as for the
standard cosmological SGRBs \citep{Goldstein2017,Savchenko2017}.  A
massive broadband follow-up campaign to GW170817 allowed an
unprecedented data coverage, revealing a kilonova/macronova (e.g.,
\citealt{Arcavi2017,Coulter2017,Cowperthwaite2017,Smartt2017,Soares-Santos2017}),
as well as the signatures of a relativistic jet
\citep{Alexander2017,Alexander2018,Margutti2017,Margutti2018,Lazzati2018,Mooley2018,Ghirlanda2019}.
The mass of the ejecta was inferred to be $\approx 0.065\,M_\odot$ by
\citet{Kasen2017} and between 0.042 and 0.077\,$M_\odot$ by
\citet{Perego2017} from modeling of the the kilonova light curve. The
former analysis assumed two components, while the latter assumed three
components, each having different density, velocity, and opening
angle.  As pointed out by \citet{Barnes2016}, key uncertainties in
kilonova modeling include the emission profiles of the radioactive
decay products -- non-thermal $\beta$ particles, $\alpha$ particles,
fission fragments, and $\gamma$ rays -- and the efficiency with which
their kinetic energy is absorbed by the ejecta.

The ejecta mass is an important quantity to measure in BNS mergers: it
contains information on the equation of state (EoS) of the NS
(e.g. \citet{Hotokezaka2013}), and is a favorable site for the
production of the heaviest elements in the Universe via the
$r$-process \citep{Lattimer1974}.  As LIGO and Virgo have begun a new
observing run at higher sensitivity, more NS-NS events will be
expected, and hence any additional diagnostics of the ejecta mass will
be extremely valuable.

The role of the ejecta in determining the properties of the jet has
been studied under different angles by a number of investigators,
especially within the context of explaining the multi-wavelength
observations of GRB170817A
(e.g. \citealt{Murguia-Berthier2014,Nagakura2014,Lazzati2017a,Lazzati2018,Xie2018,Bromberg2018,Vaneerten2018,
  Lamb2018,Wu2018, Gottlieb2018,
  Beniamini2019a,Beniamini2019b,Gottlieb2019, Kathirgamaraju2019,
  Geng2019}). These data were found to be incompatible with a top-hat
jet, but rather required a 'structured jet', as produced by the
interaction of a relativistic outflow with the ejecta generated during
the merger process.

The role of the ejecta was also recently discussed
\citep{Barbieri2019} in the context of NS-BH (black hole) mergers,
together with the role of the mass and spin of the BH, in determining
the observable electromagnetic signatures accompanying the
coalescence. Additionally, the interaction of the ejecta themselves
with the surrounding interstellar medium is expected to produce a
long-lasting radio flare \citep{Hotokezaka2015}.

Given the importance of the dynamical ejecta in molding the properties
of the relativistic outflow that propagates in it, in this paper we
perform a detailed study aimed at connecting the properties of this
matter (and in particular its mass and velocity), with the jet-cocoon
structure that develops in it, and the resulting afterglow emission in
representative bands.  The sensitivity of this radiation to the ejecta
properties (as will be shown here) will provide an additional element
linking observable quantities to fundamental physics.

Our paper is organized as follows: We begin (Section~2) by deriving
analytic expressions for the structured-jet properties, connecting the
jet morphology to the properties of the merger ejecta in which it
propagates. Aided by the results of hydrodynamic (HD) numerical
simulations, we derive energy and Lorentz factor profiles as a
function of the off-axis angle.  We feed these profiles to an external
shock synchrotron code and compute the corresponding afterglow light
curves (Section~3).  Finally, we discuss the connection between this
observable quantities and the NS EoS (Sec.~4).  We summarize and
conclude in Sec.~5.

\section{The structured jet model}

In this section we discuss an analytic framework to compute the
properties of the structured jet from a binary NS merger, within the
framework of the Kompaneets approximation (e.g.,
\citealt{Begelman1989,Matzner2003,Lazzati2005,Bromberg2011,Bromberg2014,Gill2019}).
We first discuss the input of HD simulations in setting up an analytic
profile for the energy and Lorentz factor as a function of the
off-axis angle. Subsequently we derive the scale angles and energy of
the jet and cocoon components.

\vspace{0.1in}

\textit{(a) The energy and Lorentz factor profiles}

\begin{figure}[!t]
    \centering
    \includegraphics[width=\columnwidth]{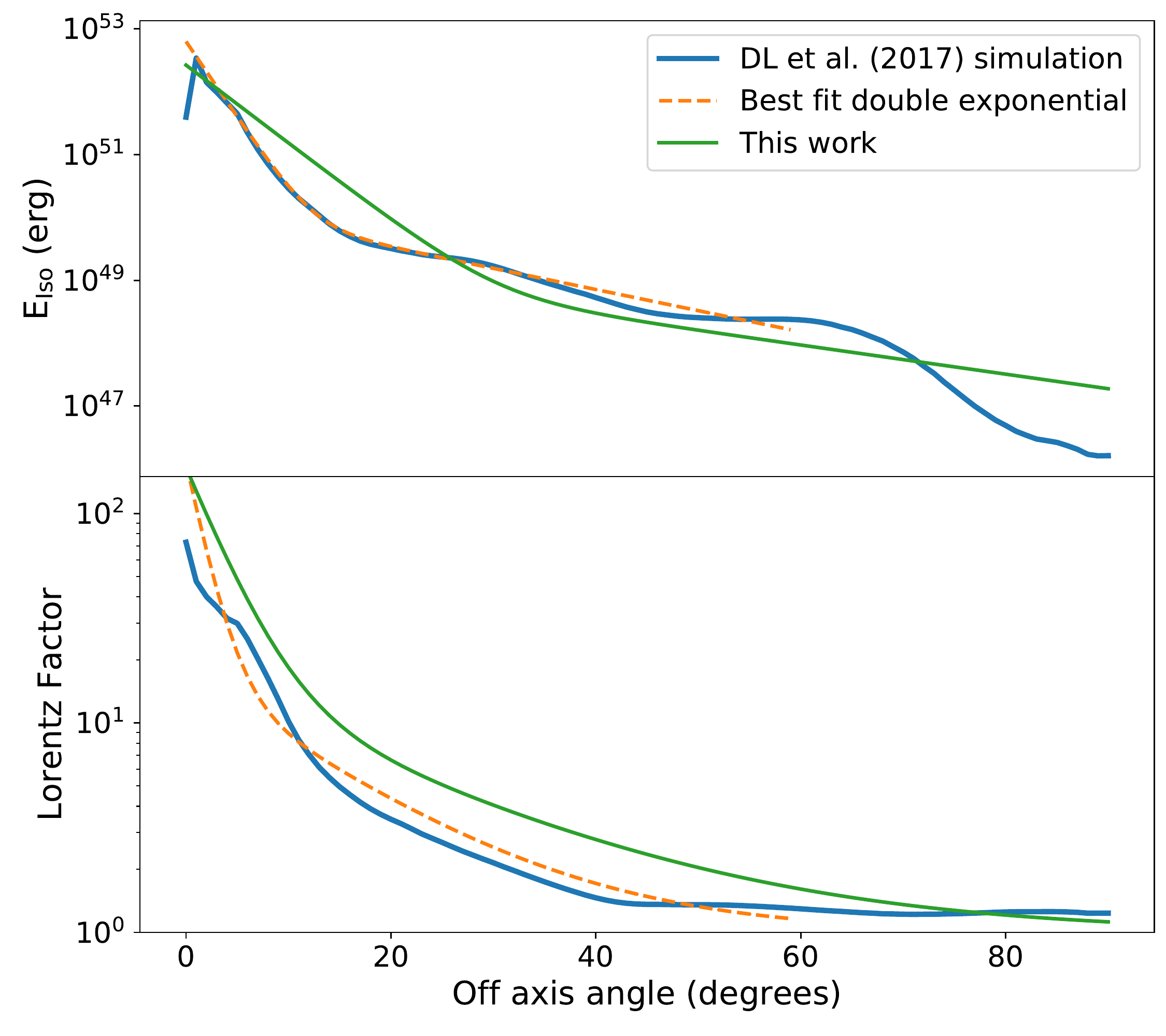}
    \caption{Comparison among the energy and Lorentz factor profile
      from a numerical HD simulation (Lazzati et al. 2017), the double
      exponential best fit function that inspired our model (see
      Eq.~\ref{eq:dblexp} and~\ref{eq:gamma}) and the profiles
      predicted by our model.}
    \label{fig:comparison}
\end{figure}

To describe the angular distribution of the relativistic outflow at
large radii (well after the jet interaction with the NR ejecta) we
adopt an exponential model, as supported by the results of numerical
simulations (Fig.~\ref{fig:comparison}):
\begin{equation}
    \frac{dE}{d\Omega}=A e^{-\frac{\theta}{\theta_j}}+B e^{-\frac{\theta}{\theta_c}}\,.
    \label{eq:dblexp}
\end{equation}
The constants $A$ and $B$ are determined by the conditions that
$A\int_\Omega e^{-\frac{\theta}{\theta_j}} d\Omega$ is the total
energy in the relativistic jet and
$B\int_\Omega e^{-\frac{\theta}{\theta_c}} d\Omega$ is the total
energy in the cocoon. The variables $\theta_j$ and $\theta_c$ indicate
the scale angles for the jet and cocoon, respectively. Note that
$\theta_j$ is in general smaller than the injected jet opening angle
$\theta_{j,\rm{inj}}$ due to HD recollimation.

A similar profile is adopted for the Lorentz factor, but the 
value is not allowed to be less than unity:
\begin{equation}
\Gamma=\Gamma_{\rm{core}} e^{-\frac{\theta}{\theta_j}}+\Gamma_{\rm{c}} e^{-\frac{\theta}{\theta_c}} +1\,.
\label{eq:gamma}
\end{equation}
where $\Gamma_{\rm{core}}$ is a free parameter of the model. The
cocoon maximum Lorentz factor $\Gamma_{\rm{c}}$ is instead the result
of the complex interaction between the jet and the ejecta, and is
mainly controlled by the mixing between the two components. This
cannot be modeled analytically and we assume here that
$\Gamma_{\rm{c}}=\Gamma_{\rm{core}}/10$, a value that fits well our
numerical results (see the bottom panel of
Figure~\ref{fig:comparison}). Given the fact that we only have one
simulation to compare to, this value should be considered
tentative. It should be remembered, however, that the initial Lorentz
factor does not affect significantly the afterglow light curve at
times longer than the deceleration time.

We now focus on calculating the energy profile (top panel of
Figure~\ref{fig:comparison}). To evaluate the values of the four
structure constants $A, B, \theta_j$, and $\theta_c$we model the
propagation of a jet through the non-relativistic (NR) ejecta. We
adopt the Kompaneets approximation
\citep{Begelman1989,Matzner2003,Lazzati2005,Bromberg2011,Bromberg2014}
and identify the scale angles $\theta_j$ and $\theta_c$ as the opening
angles of the jet and cocoon at break-out.

\vspace{0.1in}

\textit{(b) The NS ejecta}

The ejecta structure described above is created by the interaction of
a relativistic jet with non-relativistic ejecta from the merger. We
model the ejecta as a non-relativistic wind with constant velocity
$v_{\rm{w}}$ and mass loss rate $\dot{m}_{\rm{w}}$ that is released
during the merger and by the remnant a time $\delta t$ before the jet
is launched.  The ejecta are assumed to be spherically symmetric
within a solid angle $\Omega_{\rm{w}}$. The density of the ejecta is
therefore given by:
\begin{equation}
\rho_{\rm{w}}=\frac
{\dot{m}_{\rm{w}}}
{\Omega_{\rm{w}}r^2v_{\rm{w}}}\,.
\label{eq:rhow}
\end{equation}

\vspace{0.1in}

\textit{(c) The jet}

At time $\delta{t}$ after the NR ejecta are launched, a jet is
launched from the central engine. We consider a top-hat jet of
luminosity $L_{\rm{j}}$ with uniform injection Lorentz factor
$\Gamma_{\rm{j,inj}}$ within an injection opening angle
$\theta_{\rm{j,inj}}$. The jet is injected very simple, its
"structure" is acquired during the interaction with the ejecta.

One can qualitatively understand the jet propagation as follows. The
jet head propagates through the ejecta driving a bow shock that
inflates a cocoon of hot material that enshrouds the jet. The cocoon
pressure acts on the jet, collimating it into a narrower angle.  The
cocoon is trapped as well inside the ejecta and drives a shock into
them.  The model has three fundamental unknowns: the velocity of the
head of the jet $v_{\rm{jh}}$, the opening angle of the jet at
breakout $\theta_{\rm{j}}$, and the opening angle of the cocoon at
breakout $\theta_{\rm{c}}$. The system can be solved because it is
constrained by three pressure balances as will be discussed in the
part \textit{(d)} below.

The jet propagates inside the ejecta and eventually breaks out when
the outer radius of the ejecta coincides with the position of the jet
head, i.e. when
\begin{equation}
    v_{\rm{w}}\left(t_{\rm{bo}}+\delta{t}\right)=v_{\rm{jh}}t_{\rm{bo}}\,.
\end{equation}
This yields a break out time (measured from the launching of the jet) 
\begin{equation}
    t_{\rm{bo}}=\frac{v_{\rm{w}}}{v_{\rm{jh}}-v_{\rm{w}}} \delta{t}\,,
\end{equation}
and a breakout radius \citep{Murguia-Berthier2014}
\begin{equation}
    r_{\rm{bo}}=v_{\rm{jh}}t_{\rm{bo}}=
    \frac{v_{\rm{jh}}v_{\rm{w}}}{v_{\rm{jh}}-v_{\rm{w}}}\delta{t}\,.
    \label{eq:rbo}
\end{equation}

\vspace{0.1in}

\textit{(d) The pressure balances}

The structures of the jet and cocoon are defined by three pressure
balances: at the head of the jet, the ram pressure of the jet on the
contact discontinuity ($p_{\rm{ram,j}}$) is balanced by the ram
pressure of the merger ejecta on the same contact discontinuity
($p_{\rm{ram,wj}}$). At the jet-cocoon transition, the cocoon thermal
pressure ($p_{\rm{c}}$) is balanced by the jet pressure ($p_{\rm{j}}$,
be either the jet internal pressure or the ram pressure due to the
deflection of the jet material). Finally, at the cocoon-ejecta
transition, the thermal pressure of the cocoon is balanced by the ram
pressure of the merger ejecta ($p_{\rm{ram,wc}}$). These conditions
yield the equations:
\begin{eqnarray}
    p_{\rm{ram,j}}&=&p_{\rm{ram,wj}} \label{eq:pbalance1}\,, \\
    p_{\rm{j}}&=&p_{\rm{c}}\,. \label{eq:pbalance3}\,, \\
    p_{\rm{c}}&=&p_{\rm{ram,wc}}\label{eq:pbalance2}
\end{eqnarray}
This set of equation allows us to solve for the three unknowns,
$v_{\rm{jh}}$, $\theta_{\rm{j}}$, and $\theta_{\rm{c}}$.

The first pressure balance (Eq.~\ref{eq:pbalance1}) has been studied
in detail previously
(e.g. \citealt{Marti1994,Matzner2003,Lazzati2005,Bromberg2011}) and
yields the speed with which the jet head advances in the merger
ejecta. In static ejecta, the velocity of the jet head reads
\begin{equation}
    v^\prime_{\rm{jh}}=\frac{c}{1+\sqrt{\frac{\rho_{\rm{w}}}{4p_{\rm{j}}\Gamma_{\rm{j}}^2}}}\,,
    \label{eq:vjhprime}
\end{equation}
where we have made the assumption that the jet material moves with
$v_{\rm{j}}\approx{c}$ and that the enthalpy density is dominated by
the pressure component.

If the jet propagates in a moving medium, the velocity in
Eq.~(\ref{eq:vjhprime}) needs to be relativistically added to the wind
velocity to yield the jet head velocity
\citep{Murguia-Berthier2017,Matsumoto2018,Gill2019}
\begin{equation}
    v_{\rm{jh}}=\frac{v_{\rm{w}}+v^\prime_{\rm{jh}}}
    {1+\frac{v_{\rm{w}}v^\prime_{\rm{jh}}}{c^2}}\,.
    \label{eq:vjh}
\end{equation}

\begin{figure*}[!t]
\centerline{\includegraphics[width=\textwidth]{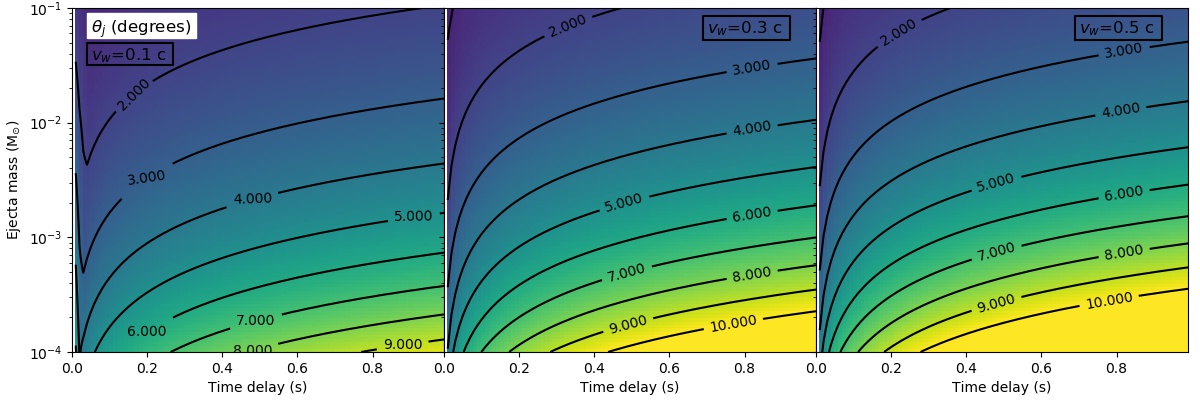}}
\centerline{\includegraphics[width=\textwidth]{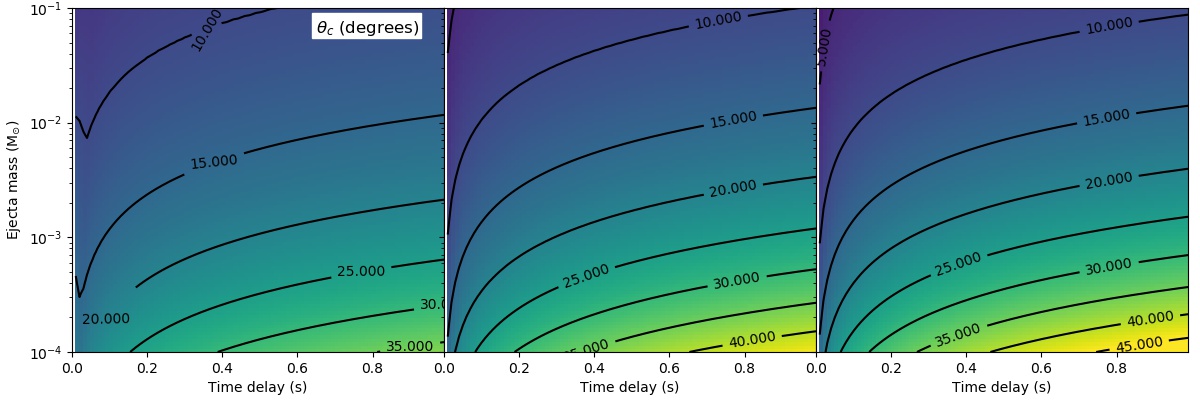}}
\centerline{\includegraphics[width=\textwidth]{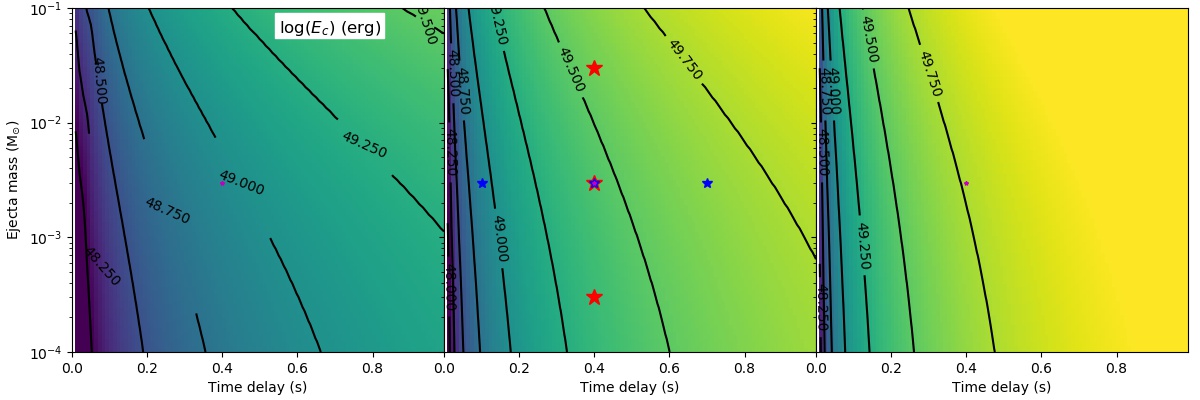}}
\caption{Jet-cocoon properties as a function of the total ejected mass
  at the time of the jet launching ($\dot{m}_{\rm{w}}\,\delta{t}$) and
  of the time delay between the release of the ejecta and of the
  relativistic jet. From top to bottom, the figures show contours of
  the jet scale angle $\theta_{\rm{j}}$, the cocoon scale angle
  $\theta_{\rm{c}}$ and of the total energy in the cocoon
  $E_{\rm{c}}$. From left to right, figures show the cases of ejecta
  velocity $v_{\rm{w}}=0.1\,c$, $0.3\,c$, and $0.5\,c$. The red stars
  in the bottom central panel show the values for which energy
  profiles are shown in Figure~\ref{fig:profiles}. The jet parameters
  are: $L_{\rm{j}}=10^{50}$~erg/s, $\theta_{\rm{j,inj}}=10^\circ$,
  $r_0=10^7$~cm, $\Gamma_0=1$. The wind is assumed to be isotropic
  ($\Omega_{\rm{w}}=4\pi$).}
\label{fig:params}
\end{figure*}

\begin{figure*}[!t]
    \centering
    \includegraphics[width=\textwidth]{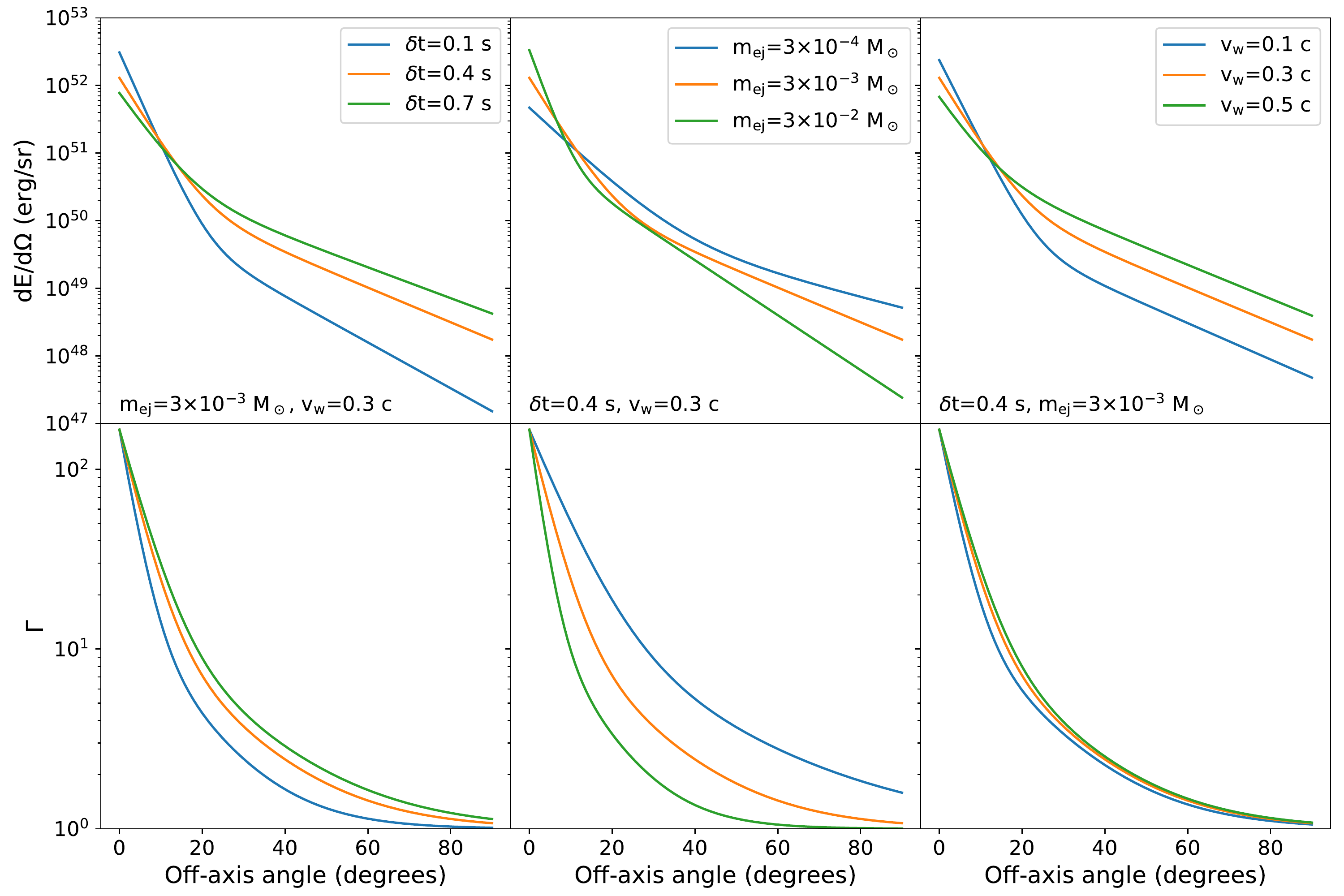}
    \caption{Energy profiles (top panels) and  Lorentz factor profiles (bottom panels) for the cases indicated with colored star symbols in Figure~\ref{fig:params}
    (see also figure legend).}
    \label{fig:profiles}
\end{figure*}

Eq.~(\ref{eq:vjhprime}) can be written in terms of unknown quantities
and model parameters. The wind density is given by Eq.~(\ref{eq:rhow})
and, using
$4p_{\rm{j}}\Gamma_{\rm{j}}^2r^2=L_{\rm{j}}/(\Omega_{\rm{j}}c)$, we
obtain:
\begin{equation}
    v^\prime_{\rm{jh}}=\frac{c}
    {1+\sqrt{\frac{\dot{m}_{\rm{w}}\Omega_{\rm{j}}c^3}
    {v_{\rm{w}}\Omega_{\rm{w}}L_{\rm{j}}}}}\,.
    \label{eq:vjh2}
\end{equation}
For the second pressure balance, we assume that the cocoon energy is
dominated by radiation so that $p_{\rm{c}}=E_{\rm{c}}/(3V_{\rm{c}})$,
where $V_{\rm{c}}=\Omega_{\rm{c}}r^3$ is the cocoon volume and
\begin{equation}
    E_{\rm{c}}=L_{\rm{j}} \left(t_{\rm{bo}}-\frac{r_{\rm{bo}}}{c}\right) \sim L_{\rm{j}} t_{\rm{bo}}
    \label{eq:ec}
\end{equation}
is the cocoon energy. Combining the above, we obtain an expression for the cocoon pressure,
\begin{equation}
    p_{\rm{c}}=\frac{L_{\rm{j}} v_{\rm{w}} \delta{t}}
    {3 r^3 \Omega_{\rm{c}} \left(v_{\rm{jh}}-v_{\rm{w}}\right)}\,.
    \label{eq:pcocoon}
\end{equation}
The ram pressure of the wind material on the cocoon is given by $p_{\rm{ram,wc}}=\rho_{\rm{w}}v_{\perp \rm{c}}^2$, where $v_{\perp \rm{c}}^2=\Omega_{\rm{c}} r^2/(\pi \delta{t}^2)$ is the velocity squared of the shock front driven by the cocoon into the ejecta. We obtain:
\begin{equation}
    p_{\rm{ram,wc}}=\frac{\dot{m}_{\rm{w}} \Omega_{\rm{c}} 
    \left( v_{\rm{jh}} - v_{\rm{w}}\right)^2}
    {\pi \Omega_{\rm{w}} v_{\rm{w}}^3 \delta{t}^2}\,.
    \label{eq:pramwc}
\end{equation}
Finally, let us consider the jet pressure, which has two components. First, the internal pressure of the jet, which is a relativistic invariant, and given by:
\begin{equation}
    p_{\rm{j,internal}}=\frac{L_{\rm{j}}\Omega_{\rm{j,inj}}r_0^2}
    {4\Omega_{\rm{j}}^2r^4\Gamma_{\rm{j,inj}}^2c}\,.
    \label{eq:pj}
\end{equation}
Additionally, one must consider the ram pressure of the jet material that is deflected from its initially radial velocity into a cylindrical flow in which the velocity is predominantly axial. This pressure component is obtained by correcting the jet ram pressure with a geometrical factor $\sin^2\theta_{\rm{j,inj}}$, yielding
\begin{equation}
    p_{\rm{ram,jc}}=\frac{L_{\rm{j}}}{\Omega_{\rm{j}} c r^2} \sin^2\theta_{\rm{j,inj}}\,.
    \label{eq:pramj}
\end{equation}

In most cases, unless the jet is not significantly recollimated, the ram pressure dominates and hence we will use Eq.~(\ref{eq:pramj}) for the jet pressure on the cocoon hereafter.

Using Eq.~\ref{eq:pcocoon} and~\ref{eq:pramwc} in Eq.~\ref{eq:pbalance2} we obtain an equation for the cocoon solid angle as
\begin{equation}
    \Omega_{\rm{c}}^2=\frac{\pi L_{\rm{j}} v_{\rm{w}}^4 \delta{t}^3 \Omega_{\rm{w}}}
    {3 r^3 \left(v_{\rm{jh}}-v_{\rm{w}}\right)^3 \dot{m}_{\rm{w}}}
    \label{eq:omc1}\,.
\end{equation}
Alternatively, an expression for the cocoon solid angle can be derived
using Eq.~\ref{eq:pcocoon} and~\ref{eq:pramj} in Eq.~\ref{eq:pbalance3} to yield
\begin{equation}
    \Omega_{\rm{c}}=\frac{\Omega_{\rm{j}} c v_{\rm{w}} \delta{t}}
    {3 r \left(v_{\rm{jh}}-v_{\rm{w}}\right) \sin^2\theta_{\rm{j,inj}}}\,.
    \label{eq:omc2}
\end{equation}

Equating Eq.~(\ref{eq:omc1}) to the square of~(\ref{eq:omc2}) and solving for the jet solid angle we obtain:
\begin{equation}
    \Omega_{\rm{j}}=\sqrt{\frac{3\pi L_{\rm{j}} v_{\rm{w}} \Omega_{\rm{w}}}
    {\dot{m}_{\rm{w}} c^2 v_{\rm{jh}}}}\sin^2\theta_{\rm{j,inj}}\,.
\end{equation}

The right hand side of this equation can now be inserted for the jet opening angle in Eq.~(\ref{eq:vjh2}) and solved for the jet head velocity, by using Eq.~(\ref{eq:rbo}) for the radius $r$. 

Finally, the cocoon energy can be derived from
Equation~\ref{eq:ec}
and, thanks to energy conservation, the remaining jet energy $E_{\rm{j}}=E_{\rm{j,inj}}-E_{\rm{c}}$, where $E_{\rm{j,inj}}$ is the energy injected at the base of the jet.

\begin{figure*}[!t]
    \centering
    \includegraphics[width=\textwidth]{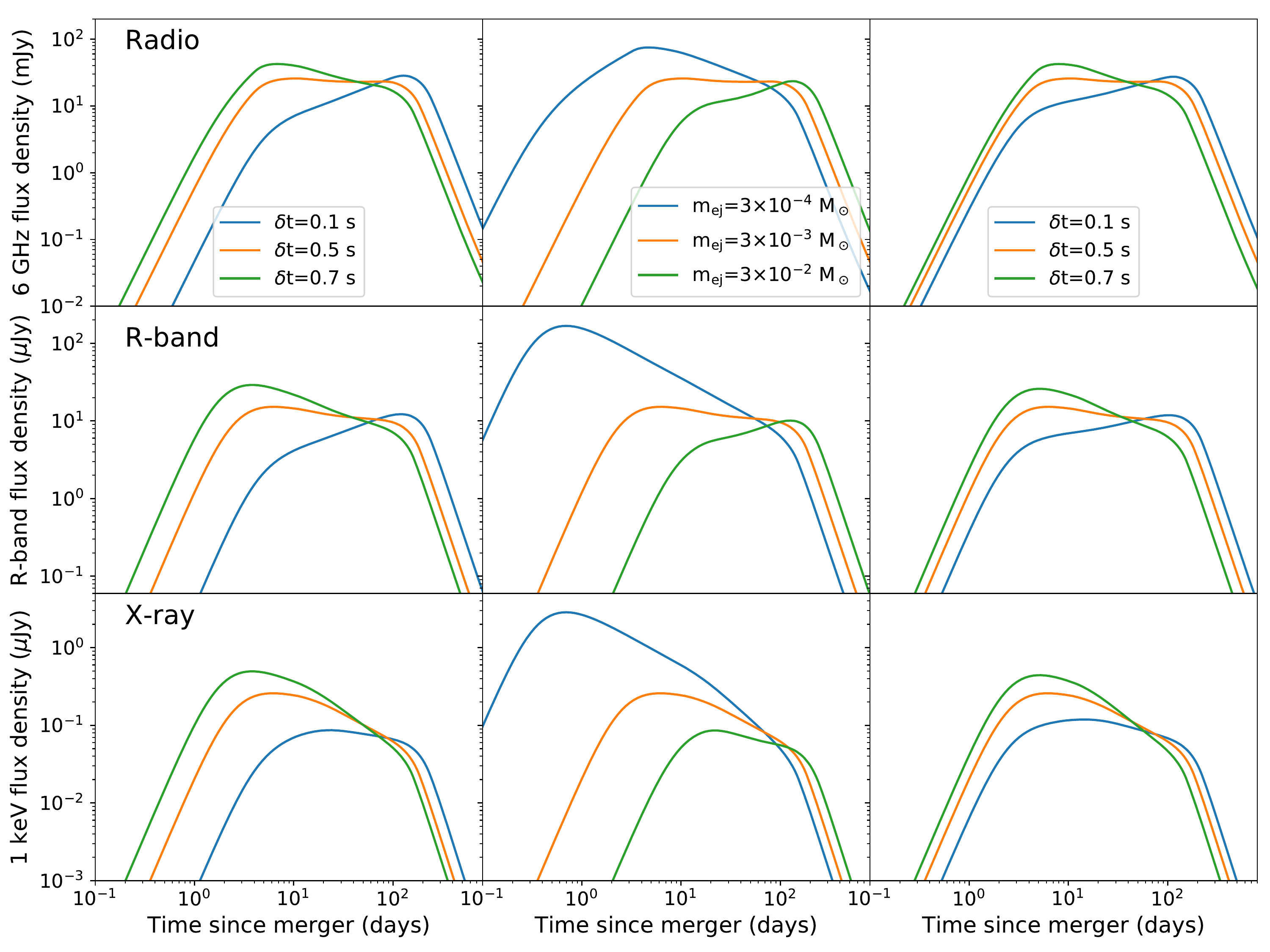}
    \caption{X-ray (top panels), optical (middle panels), and Radio light curves (bottom panels) for the cases indicated with colored star symbols in Figure~\ref{fig:params}. Light curves were calculated for a uniform interstellar medium with density $n_{\rm{ISM}}=10^{-2}$~cm$^{-3}$ for an observer $30^\circ$ away from the jet axis. The shock microphysics parameters were set to $\epsilon_{\rm{e}}=0.1$, $\epsilon_{\rm{B}}=0.01$, and $p_{\rm{el}}=2.3$. The distance to the burst was set to 40 Mpc. }
    \label{fig:lcurves}
\end{figure*}
    
To validate this model we show in Figure~\ref{fig:comparison} a
comparison between the numerical results of Lazzati et al. (2017) and
the energy and gamma profiles derived from the above equations. A
double exponential fit (orange dashed line) to the numerical results
(blue solid line) yields $\theta_{\rm{j}}=1.8^\circ$ and
$\theta_{\rm{c}}=12.9^\circ$, respectively. Our analytic model (green
solid line) yields instead $\theta_{\rm{j}}=3.5^\circ$ and
$\theta_{\rm{c}}=18.7^\circ$. While there are some differences due to
the complexity of the interaction between a relativistic jet and the
relatively slow and dense ejecta, the main features are well
represented in the analytic model. We stress that the simulation was
used as an input for the functional form in Equations~\ref{eq:dblexp}
and~\ref{eq:gamma}, but the scale angles and cocoon energy used to
draw the green curve in the figure are obtained form the analytic
model. The general agreement between the numerical result (in blue)
and the model (in green) is therefore not a circular argument.

Figure~\ref{fig:params} shows the jet and cocoon opening angles (top
and middle panels), and cocoon energies (bottom panels) as a function
of the total mass in the ejecta at the time of the jet injection and
of the delay between the launching of the ejecta and the release of
the relativistic jets. From left to right, ejecta velocities
$v_{\rm{w}}=0.1$~$c$, 0.3~$c$, and 0.5~$c$ are shown.

Figure~\ref{fig:profiles} shows the energy (top panels) and Lorentz
factor (bottom panels) profiles for a few selected cases. The cases
shown in Figure~\ref{fig:profiles} are indicated in
Figure~\ref{fig:params} with colored star symbols.

\section{Afterglow light curves}

We compute the afterglow light curves by feeding the energy and
Lorentz factor profiles into a semi-analytic external shock code. We
adopt the same version of the radiative code that was previously used
by \citet{Lazzati2017a,Lazzati2017b, Lazzati2018} and
\citet{Perna2019}.  The code uses an analytic approximation for the
$r(t)$ profile that correctly approaches asymptotically the
relativistic limit ($r\propto{t}^{1/4}$) at early times and the
Sedov-Taylor scaling ($r\propto{t}^{2/5}$) at late times. However,
throughout the evolution the code assumes a Blandford-McKee
self-similar profile \citep{Blandford1976} for the blast wave
structure behind the shock. This yields an approximate evolution for
the afterglow, especially at the later stages, when the transition to
a non-relativistic expansion alters the downstream shock profile
(e.g., \citealt{Huang1999,Huang2000,DeColle2012,Peer2012,Li2019}). The
light curves shown here, when compared with more accurate calculations
such as those used in BoxFit (\citealt{vanEerten2012}), are correct
within $\sim20\%$ even at late times.

The external material is assumed to be a low-density, uniform
interstellar medium of density $n_{\rm{ISM}}$, and the shock
microphysics is parameterized, as customary, with the equipartition
parameters $\epsilon_{\rm{e}}$ and $\epsilon_{\rm{B}}$ and with the
non-thermal electron's energy distribution index $p_{\rm{el}}$. All
electrons are assumed to be accelerated into the non-thermal
distribution.  We perform our light curves calculations adopting
typical average values for the shock microphysics parameters, as
inferred in the afterglow modeling of the \textit{Swift}-detected
short GRBs \citep{Fong2015} and of GRB170817
\citep{Lazzati2017b}. These are: $\epsilon_{\rm{e}}=0.1$,
$\epsilon_{\rm{B}}=0.01$, and $p_{\rm{el}}=2.3$.  The interstellar
medium is assumed to be uniform and tenuous, as expected in the
surrounding of a BNS merger, with density
$n_{\rm{ISM}}=10^{-2}$~cm$^{-3}$.  The observer angle with respect to
the jet axis is set to $\theta_{\rm jet}=30^\circ$, the typical
viewing angle expected for Ligo/Virgo detected events
\citep{Schutz2011} and the distance to the burst is set to 40 Mpc as a
representative case to relate to GRB170817.

Examples of light curves in X-ray (1 keV), optical (R-band), and radio
(6 GHz) are shown in the top and bottom panels of
Figure~\ref{fig:lcurves}, respectively. Interestingly, the light
curves are noticeably different from one another, giving hope that the
ejecta mass and velocity, as well as the time delay $\delta{t}$, can
be constrained with a high quality dataset. In particular, the slope
of the slow-rising phase before the maximum and the prominence of the
feature at the peak are characteristic of the jet-cocoon
configuration. A larger ejecta mass results in a more energetic cocoon
confined in a narrower scale angle $\theta_{\rm{c}}$
(Figure~\ref{fig:params}) that moves with lower Lorentz factor
(Figure~\ref{fig:profiles}). The lower Lorentz factors cause a late
onset of the afterglow at both X-ray and radio frequencies, which is
therefore a signature of significant mass ejection. Large ejecta
masses also cause the jet scale angle $\theta_{\rm{j}}$ to narrow
significantly, resulting in a clearer jet feature at the light curve
peak, especially at radio frequencies (see, e.g., the green light
curve in the top-center panel of Figure~\ref{fig:lcurves}). Very low
mass ejecta produce instead a very weak but fast moving cocoon and
wide angle jet. These light curves peak early, especially in X-rays,
and more closely resemble those of canonical, on-axis jets. Note that
in the complete absence of ejecta, the light curves peak late, since
no cocoon is present, as shown with the case of a pure jet by
\citet{Perna2019}. This might seem contradictory, but there is indeed
a big difference between the complete absence of ejecta, which leaves
the top-hat jet unaltered, and the presence of light ejecta, which
produce a fast, low-energy cocoon. The former configuration gives a
peak time that corresponds with the time at which the jet emission
comes into the line of sight, at about $100$~s in
Figure~\ref{fig:lcurves}. The latter, instead, peaks early, since the
deceleration radius scales as
$r_{\rm{dec}}\propto{E_c}^{1/3}\Gamma_{\rm{c}}^{-2/3}$, yielding a
deceleration time for material along the line of sight
$t_{\rm{dec}}\propto{E_c}^{1/3}\Gamma_{\rm{c}}^{-8/3}$. This yields a
scaling for the cocoon-afterglow peak luminosity
$L_{\rm{pk}}\propto{E_{\rm{c}}}^{2/3}\Gamma_{\rm{c}}^{8/3}$, showing
that the peak luminosity of the cocoon grows quickly for increasing
$\Gamma_{\rm{c}}$, unless its energy vanishes.

In some cases, however, there are some degeneracies. For example, a
careful comparison of the leftmost and rightmost panels of
Figure~\ref{fig:lcurves} reveals similar light curves at both radio
and X-ray frequencies. This is not surprising since both the delay
$\delta{t}$ and the ejecta velocity $v_{\rm{w}}$ control the radius of
the ejecta in which the jet is propagating. Breaking these
degeneracies will require the use of additional data. For example, one
could use the shock-breakout model of \citet{Kasliwal2017} to
constrain the outer radius of the ejecta from the brightness of the
prompt $\gamma$-ray emission. In addition, constraints on $\delta{t}$
can be set since it has to be smaller than the time delay between the
detection of gravitational waves and of $\gamma$-ray
photons. Additional constraints could come form the kilonova
properties and from the gravitational wave detection (e.g.,
constraints on the inclination of the system, the chirp mass, the
individual masses of the two NS). Thus one can envisage that afterglow
modeling with the appropriate priors from the complete dataset will be
able to set meaningful constraints on the ejecta properties.

\begin{figure*}[!t]
    \centering
   \includegraphics[width=\textwidth]{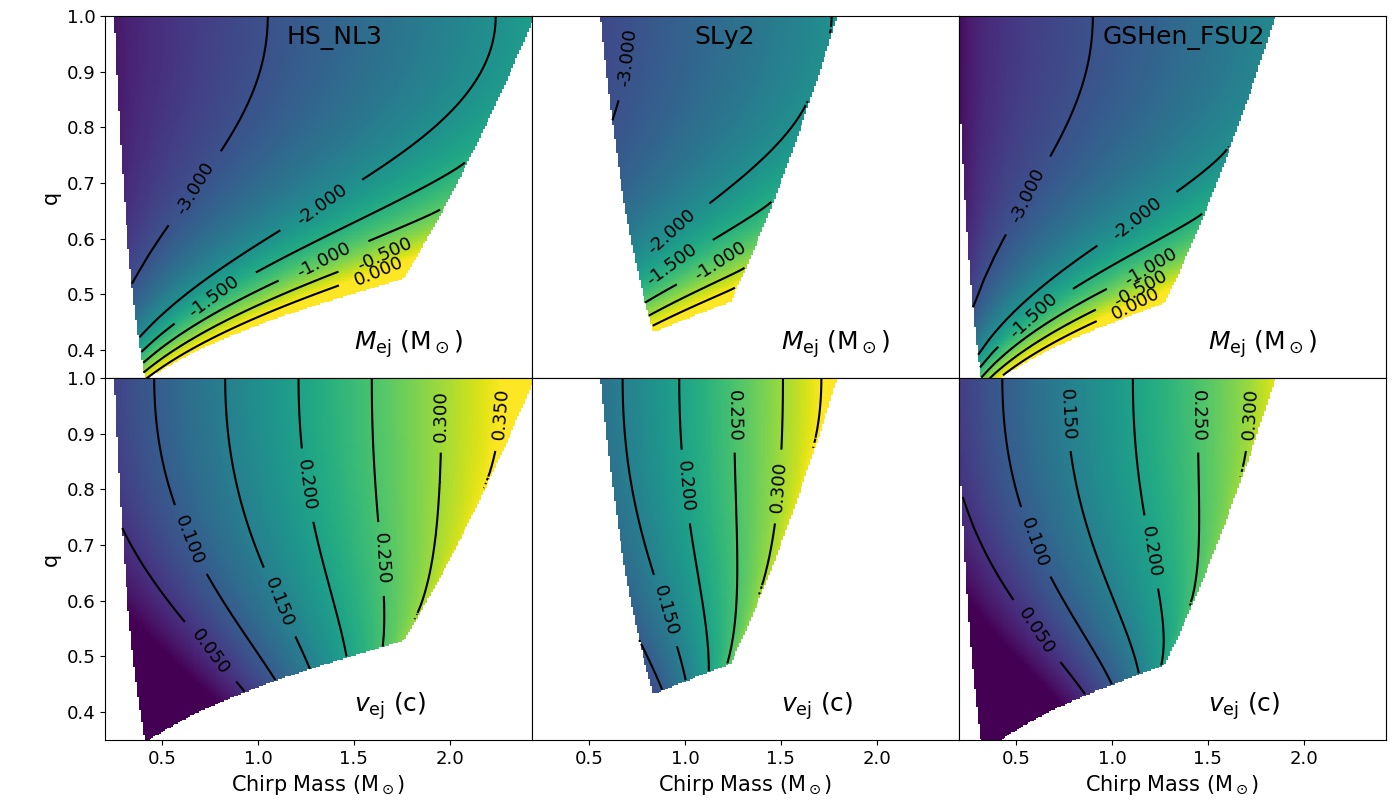}
   \caption{Ejecta mass (top panels) and ejecta velocity (bottom
     panels) produced during a binary merger, for three representative
     equations of state of neutron star matter. The plots are made
     using the fits  by \citet{Coughlin2018a. These figures are
       only representative here since they display the dynamical
       component of the ejecta (see text for discussion). }}
    \label{fig:eos}
\end{figure*}

\section{Ejecta properties and the Equation of state of neutron stars}

As already mentioned in \S1 to motivate this work, the connection
between the amount and velocity of the ejected mass and the observable
properties of the mergers (in particular in the electromagnetic
spectrum) is especially interesting in light of the fact that it
contains information on the EoS of the NS, as demonstrated by several
groups via GRMHD simulations
(e.g. \citealt{Shibata2005,Kiuchi2010,Rezzolla2010,Hotokezaka2011,
  Bauswein2013,Hotokezaka2013,Giacomazzo2013b,
  Sekiguchi2016,Lehner2016,Ruiz2016,Kawamura2016,Radice2016,Ciolfi2017,
  Dietrich2017a,Radice2018,Foucart2019})\footnote{ Recent work
  \citep{Most2019} has also highlighted the impact of the NS spins on
  the ejecta mass.}.

Most of the tidally disrupted mass remains bound and circularizes into
a hot torus, whose accretion onto the remnant compact object formed
from the merger powers a relativistic jet. The total energy in the
jet, proportional to the total accreted mass, carries information on
the EoS; this connection was exploited by \citet{Giacomazzo2013a} with
the sample of the short GRBs available at the time, to draw inferences
on the likelihood of various EoS.

Here we are focusing on signatures of the ejecta mass in the light
curves.  This high-velocity material is made of several
components. Firstly, the dynamical ejecta
(e.g. \citealt{Bauswein2013,Hotokezaka2013,Sekiguchi2016,Lehner2016,Dietrich2017a,Radice2018}),
which is partly due to matter ejected (and unbound) by the tidal
disruption of the less massive NS, and partly by matter ejected (and
unbound) during merger due to the formation of shocks.  More material
is expected to be ejected via ablation from the surface of a
Hypermassive NS (HMNS) in the case that the merger remnant passes
through such a phase (e.g.
\citealt{Dessart2009,Perego2014,Just2015,Perego2017}). Additionally,
nuclear and viscous processes in the accretion disk surrounding the
remnant compact object unbound material from the disk, creating a disk
wind
(e.g. \citealt{Metzger2009,Fujibayashi2017,Fujibayashi2018,Siegel2018,Radice2018,Fahlman2018}),
also referred to as secular ejecta. Recent simulations
(e.g. \citealt{Siegel2017,Radice2018}) find that the secular ejecta
mass typically exceed the mass of the dynamical ejecta, except for the
cases of prompt BH formation, in which they can be comparable (albeit
small). The quantitative results for the masses of both components are
still rather uncertain, as they depend on a number of microphysical
effects not all of which are fully included at this stage in the
numerical simulations. Neutrino cooling for example plays an important
role, and so does the magnitude of the effective viscosity arising
from magnetohydrodynamic instabilities operating during the
merger. For example, \citet{Radice2018b} found that the mass of the
fast tail of the dynamical ejecta can vary by up to four orders of
magnitude depending on the viscosity strength.

In terms of relevance for our problem the ejecta mass and velocity,
their launch time with respect to the jet, and their angular
distribution play a role. \citep{Fernandez2013} showed that disk
outflows can develop only after the disk becomes advective and
$\alpha$-particles form. They find that the mass flux takes about one
second before reaching its maximum. If the disk wind is in fact
delayed with respect to the jet launch, then the jet will mostly
interact with the dynamical ejecta.

What matters for this work is the fact that both the dynamical ejecta
and the wind outflow carry the imprint of the EoS of the NS, in
addition to being dependent on the masses and mass ratios of the NSs.
Generally speaking, the simulations show that larger mass ratios tend
to result in the partial disruption of the smaller star during the
merger, and hence a smaller amount of shocked material, since the
stars merge less violently. The mass asymmetry also produces more
massive tidal outflows. A larger amount of ejecta further tends to
correlate with softer EoSs.

A visual representation of the dependence of the ejecta properties on
the EOS of NS is offered in Figure~\ref{fig:eos}, which displays the
dynamical ejecta mass (top panels) and velocity (bottom panels) for
three representative equations of state of neutron star
matter\footnote{Data from https://compose.obspm.fr/}. These quantities
are shown as a function of the mass ratio $q$ and the chirp
mass, which is extremely well constrained by GW observations alone
(e.g., \cite{Abbott2018}). In the figures, white areas correspond to
regions that are either unphysical or for which the mass-radius
relationship is not allowed by that equation of state.  The plots of
Figure~\ref{fig:eos} were made using the fits by
\citet{Coughlin2018a}, based on an update of the ones by
\citet{Dietrich2017} with the inclusion of more simulations (for a
total of 259); however they describe only the dynamical component of
the ejecta, for which a much larger number of simulations (and
parameter exploration) exists in the literature to allow the
derivation of fitting formulae.  For the disk wind and the HMNS wind
component fewer simulations exist. Some fits to the ejecta mass have
been provided in the literature
(e.g. \citealt{Radice2018,Coughlin2018b}), but no systematic fit
exists to date for the ejecta speed (see however some correlations
with the disk mass provided by \citealt{Fahlman2018}). Therefore, the
connection between ejecta properties and EoS will need to be further
tightened in the future with more accurate simulations. At the same
time, it will also be important to better understand the time of jet
launch with respect to the other outflow timescales in the problem.

\section{Summary and Conclusions}

The detection of electromagnetic radiation from a binary NS merger
triggered by GWs has allowed an unprecedented view into the inner
workings of a relativistic jet, likely powered by accretion onto the
remnant object, and propagating onto the ejecta from the merger.

Since the LIGO horizon is much smaller than the \textit{Swift} one,
events triggered in GWs will be generally much closer than the ones
triggered in $\gamma$-rays. Hence their afterglows, especially in the
radio band which is not contaminated by the kilonova emission, will
generally be much brighter than for the more distant short GRBs
triggered in $\gamma$-rays.  As shown by the case of GW170817, this
allows for a detailed monitoring of the afterglow light curves in
several bands by a variety of instruments, and hence a detailed shape
reconstruction of the light curve.

In this paper we have studied the structure of the jet-cocoon which
forms as a result of the jet expanding into the fast-moving ejecta
produced by the NS-NS merger. More specifically, we have explored how
this structure varies with the density, velocity, and radial extent of
the ejecta. We have then used this structure at the radiative stage to
compute the shape of the afterglow in the X-ray and radio bands (the
optical is likely dominated by the kilonova, hence may have less
diagnostic power), and explored its dependence on the ejecta
properties, as well as on the time delay between merger and onset of
the jet.

We identified features in the light curves which are especially
distinctive of a jet-cocoon configuration, where the cocoon is
generated by the interaction of the jet with the dynamical ejecta. A
more pronounced cocoon yields an additional, earlier bump or plateau
in the light curve, compared to the singly-peaked light curve of a
pure relativistic jet. The rise time of the light curves is also very
sensitive to the amount of ejecta, with larger values yielding delayed
rises.  It should be noted that our scenario might be oversimplified
in some aspects and further observations or advancement in the
understanding of the ejecta of BNS mergers could spark the need for
modifications. In particular, the ejecta might not be spherically (or
even cylindrically) symmetric. An oblate configuration, for example,
would cause more efficient collimation of the jet (e.g.,
\citealt{Duffell2015}). In addition, the central engine might not
become dormant after a few seconds, as assumed here. A long-active
central engine, such as a magnetar, may result in extra energy
injection in both the jet (\citealt{Geng2018}) and the ejecta
(\citealt{Yu2018}), altering the simple geometry and dynamics that we
assumed.

In conclusion, a detailed monitoring of the afterglow light curve,
especially in the cleaner X-ray and radio bands, can help constrain
the properties of the ejecta from the merger event.  Since this
material is sensitive to the equation of state of dense matter (in
addition to being important for production of heavy elements),
detailed monitoring of afterglow light curves of GW-detected binary NS
mergers (and NS-BH alike) has the power to help probing the physics of
the merger events and the equation of state of dense nuclear matter.

\vspace{0.25in}

\acknowledgments We thank Riccardo Ciolfi, Tim Dietrich, Diego
L\'opez-C\'amara, Brian Metzger, David Radice, Om Sharam Salafia, and
Eleonora Troja for fruitful discussions and constructive comments on
an early draft of this manuscript, and the referee for thoughtful
comments and suggestions. DL acknowledges support from NASA ATP grant
NNX17AK42G, NASA Fermi GI grant 80NSSC19K0330, and Chandra grant
TM9-20002X. RP acknowledges support from the NSF under grant
AST-1616157.  The Center for Computational Astrophysics at the
Flatiron Institute is supported by the Simons Foundation.

\bibliographystyle{aasjournal}
\bibliography{biblio}

\end{document}